\documentclass[aps,prl,twocolumn,preprintnumbers,amsmath,amssymb,longbibliography,superscriptaddress,a4paper,10pt,linenumbers10pt]{revtex4-2}

\usepackage{preamble}

\begin{document}

\title{Many-body dynamical localization in Fock space}

\author{N.~Dupont}\email[Corresponding author: ]{nathan.dupont@lkb.upmc.fr}
\affiliation{Center for Nonlinear Phenomena and Complex Systems, Université Libre de Bruxelles, CP 231, Campus Plaine, 1050 Brussels, Belgium}
\affiliation{International Solvay Institutes, 1050 Brussels, Belgium}
\affiliation{Laboratoire Kastler Brossel, Sorbonne Université, CNRS, ENS-Universit\'e PSL, Collège de France, 4 Place Jussieu, 75005 Paris, France}

\author{B.~Peaudecerf}
\affiliation{Laboratoire Collisions Agrégats Réactivité, Université de Toulouse, CNRS, 31062 Toulouse, France}
\author{D.~Guéry-Odelin}
\affiliation{Laboratoire Collisions Agrégats Réactivité, Université de Toulouse, CNRS, 31062 Toulouse, France}

\author{G.~Lemarié}
\affiliation{INPHYNI, Université Côte d’Azur, CNRS, Nice, France}

\author{B.~Georgeot}
\affiliation{Laboratoire de Physique Théorique, Université de Toulouse, CNRS, UPS, France}

\author{Ch.~Miniatura}
\affiliation{INPHYNI, Université Côte d’Azur, CNRS, Nice, France}
\affiliation{Centre for Quantum Technologies, National University of Singapore, Singapore}

\author{N.~Goldman}
\affiliation{Center for Nonlinear Phenomena and Complex Systems, Université Libre de Bruxelles, CP 231, Campus Plaine, 1050 Brussels, Belgium}
\affiliation{International Solvay Institutes, 1050 Brussels, Belgium}
\affiliation{Laboratoire Kastler Brossel, Coll\`ege de France, CNRS, ENS-Universit\'e PSL, Sorbonne Universit\'e, 11 Place Marcelin Berthelot, 75005 Paris, France}

\date{\today}

\begin{abstract}
We investigate the emergence of many-body dynamical localization (MBDL) in the Fock space of an interacting two-mode bosonic system subject to periodic driving. Using a mapping to the paradigmatic kicked-top model, we analyze the interplay between classical chaotic diffusion and quantum interference effects. While the mean-field (classical) dynamics exhibits bounded ergodic diffusion along the population imbalance axis, the quantum dynamics reveals strong suppression of transport in Fock space, in close analogy with Anderson localization in disordered lattices. We characterize the localization length, its scaling with particle number and driving parameters, and reveal the spectral crossover from random-matrix to Poisson statistics as the many-body ensemble localizes. 
We highlight the connection between MBDL and discrete time crystals. Our findings offer a promising avenue to study the Anderson transition in Fock space.
\end{abstract}

\maketitle

\paragraph{Introduction.} Anderson localization is a fundamental wave phenomenon whereby a particle becomes spatially localized in a disordered potential landscape~\cite{anderson_1958}. It arises from destructive interference between partial waves multiply scattered by the disorder, which suppresses the classically-expected diffusion. Crucially, Anderson localization also signifies a breakdown of ergodicity: whereas a classical particle would, in principle, explore the entire energetically accessible phase space, a quantum particle subject to localization remains confined to a finite region. Beyond static potentials, time-modulated chaotic systems can also exhibit an effective disorder, leading to \textit{dynamical} localization~\cite{casati_1979}. A prominent example is the quantum kicked rotor, in which a particle initially undergoes diffusive motion in momentum space before localizing exponentially~\cite{casati_1979, grempel_1984, moore_1995, chabe_2008, lemarie_2009}.

The question of whether localization can survive the introduction of interactions --- a phenomenon termed many-body localization (MBL) --- has attracted considerable attention over the past three decades~\cite{shepelyansky_1994,altshuler_1997,gornyi_2005,basko_2006,alet_2018,abanin_2019,sierant_2025}. Recent experiments with cold atomic gases in finite-size systems have provided compelling evidence of MBL, revealing the persistence of memory of initial conditions and the absence of thermal equilibration over experimentally accessible timescales~\cite{schreiber_2015,choi_2016}. However, the existence of a true MBL phase in the thermodynamic limit remains debated~\cite{sierant_2025}. MBL has also been studied in periodically driven systems, both theoretically and experimentally, mainly in one dimension (1D), with the interacting kicked rotor~\cite{shepelyansky_1994,borgonovi_1995,pikovsky_2008,notarnicola_2018,notarnicola_2020,lellouch_2020,russomanno_2021,vuatelet_2021,cao_2022,see_toh_2022,vuatelet_2023,guo_2025,olsen_2025,olsen_2025b,yang_2026}, driven Bose-Hubbard chains~\cite{lazarides_2015,fava_2020} and the kicked Ising model~\cite{ponte_2015,abanin_2016,lezama_2019,sierant_2023}. 

Here, we propose a minimal and experimentally relevant model exhibiting many-body dynamical localization (MBDL) in Fock space~\cite{altshuler_1997,basko_2006,de_luca_2013,bauer_2013,alet_2018,mace_2019,de_tomasi_2021,scoquart_2024,roy_2024}: an ensemble of $N$ interacting bosons in a two-mode system subject to periodic kicking of the inter-mode coupling. This system maps onto the kicked top, a paradigmatic model for quantum chaos~\cite{nakamura_1986,haake_1987,haake_2000,xie_2005}, with an effective spin of length $N/2$ setting the size of a 1D space spanned by the Fock states. In the fully chaotic classical regime ($N\rightarrow\infty$), we precisely characterize the phase-space dynamics. In the quantum regime ($N$ finite), we uncover a crossover between the extensively studied ergodic regime~\cite{peres_1994,alicki_1996,miller_1999,zyczkowski_2001,tanaka_2002,weinstein_2002,bandyopadhyay_2004,wang_2004,ghose_2004,demkowicz_2004,xie_2005,weinstein_2006,stamatiou_2007,ghose_2008,chaudhury_2009,swingle_2016,neill_2016,pappalardi_2018,piga_2019,kumari_2019,herrman_2019,lerose_2020,li_2021,wang_2023,omanakuttan_2023} and a largely unexplored regime --- hinted at by Haake and Shepelyansky~\cite{haake_1988} --- in which quantum interferences give rise to an Anderson-like dynamical localization of the many-body state. We provide an estimate for the localization length in the Fock space and identify the ergodic-to-localized crossover through spectral statistics. Finally, we establish a direct connection between MBDL and discrete time crystals (DTCs)~\cite{sacha_2015,yao_2017,choi_2017,zhang_2017,sacha_2018,zaletel_2023}, showing that DTCs can be supported by eigenstates dynamically localized near the fully polarized Fock states, and we outline how this simple two-mode system can serve as a building block for the study of MBL in larger systems.

\begin{figure}[t]
\begin{center}
\includegraphics[scale=1]{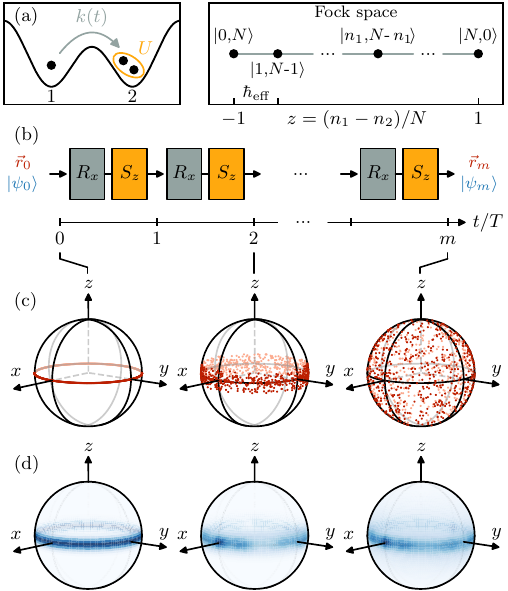}
\caption{
\textbf{Kicked system, Fock space picture and dynamical localization.} \textbf{(a)} Left: Schematics of a two-mode system~\eqref{eq:h}, with two-body interactions $U$ between bosons (black disks) and driven hopping amplitude $k(t)$. Right: Corresponding 1D Fock space.
\textbf{(b)} Depiction of the quantum (Eq.~\eqref{eq:U_kt}) and classical (Eq.~\eqref{eq:map_kt}) evolutions, with $\vec{r}_m = (x,y,z)_m$.
\textbf{(c)} Evolution of 1000 mean-field trajectories starting from the equator of the Bloch sphere, at periods $m=0$ (left), $m=2$ (middle) and $m=100$ (right), with rotation angle $a=0.1$ and torsion amplitude $b=600$.
\textbf{(d)} Average Husimi distributions~\cite{husimi_1940,agarwal_1981} of 1000 quantum states evolved by Eq.~\eqref{eq:U_kt} with $a=0.1$ and $b$ sampled from 563 to 615, at the same periods as (c), with $\ket{\psi_0}=\ket{100,100}$ ($N=200$).
}
\label{fig1}
\end{center}
\end{figure}

\paragraph{Kicked system and mean-field limit.} We consider $N$ interacting bosons in a periodically kicked two-mode system, as depicted in Fig.~\ref{fig1}~(a) for a double-well implementation in the tight-binding regime. The Hamiltonian describing the system reads
\begin{equation}
\label{eq:h}
\hH(t) \!=\! \dfrac{U}{2}(\ha_1^{\dagger2}\ha_1^2+\ha_2^{\dagger2}\ha_2^2) - \dfrac{k}{2} \!\sum_{m\in\mathbb{Z}}\delta(t-mT) (\ha_1^\dagger\ha_2 + \ha_2^\dagger\ha_1),
\end{equation}
where $\ha_\ell^\dagger$, $\ha_\ell$ are the bosonic creation and annihilation operators in mode $\ell=1,2$, $U$ is the two-body interaction energy~\footnote{Although we will consider repulsive interactions, our results do not depend on the sign of $U/k$.}, $k$ is the kick amplitude, $\delta(t)$ is the Dirac delta function and $T$ is the drive period. Through the Jordan-Schwinger mapping~\cite{schwinger_1952}, the system can be described as a quantum spin-$N/2$ $\hat{\bm{J}}$ with components
\mbox{$\hJ_x = (\ha_1^\dagger\ha_2 + \ha_2^\dagger \ha_1)/2$},
\mbox{$\hJ_y = (\ha_1^\dagger\ha_2 - \ha_2^\dagger \ha_1)/2\di$} and
\mbox{$\hJ_z = (\hn_1 - \hn_2)/2$} ($\hn_\ell$ being the number operator in mode $\ell$) satisfying the commutation relations $[\hJ_\mu,\hJ_\nu] = \di \epsilon_{\mu\nu\lambda}\hJ_\lambda$ and the invariant~$\hat{\bm{J}}^2 = \hN/2[(\hN/2)+1]$ with $\hN = \hn_1+\hn_2$. The Hamiltonian~\eqref{eq:h} is thus rewritten
\begin{equation}
    \label{eq:hkt}
    \hH(t) = U \hJ_z^2 - k \hJ_x \sum_{m\in\mathbb{Z}}\delta(t-mT) + \dfrac{\hN U}{4}(\hN-2).
\end{equation}
Since $[\hH(t),\hN] = 0$, the total particle number is conserved. Up to its last constant term, Eq.~\eqref{eq:hkt} is the Hamiltonian of the quantum kicked top~\cite{haake_1987}, whose Floquet operator over one modulation period, evaluated from just before one kick to just before the next, reads
\begin{equation}
\label{eq:U_kt}
    \hU_\dF = \de^{-\di \hbar_\deff b \hJ_z^2}\de^{-\di a \hJ_x} \equiv \hS_z(b)\hR_x(a),
\end{equation}
where $a=-k/\hbar$ and $b=NUT/2\hbar$ (with $\hbar$ the reduced Planck constant). The effective Planck constant $\heff = 2/N$ governs the classical limit --- which coincides with the mean-field limit of the many-body system --- and sets the quantization of the population imbalance $\heff \hJ_z=(\hn_1-\hn_2)/N$ (see Fig.~\ref{fig1}~(a)). In the classical limit $N\rightarrow\infty$, $\heff\rightarrow 0$, the rescaled operators $\heff \hJ_\mu$ can be approximated by the coordinates $\mu=x,y,z$ describing the state on the unit Bloch sphere~\footnote{In the mean-field limit, $U\rightarrow0$ as $N\rightarrow\infty$ such that $b$ remains constant; see~\cite{Note3}.}. This yields the classical stroboscopic map of the kicked top~\cite{haake_1987,haake_1988}:
\begin{align}
\label{eq:map_kt}
x_{m+1}\! &= \cos(b z_m) x_m - \sin(b z_m) y_m, \\
y_{m+1}\! &= \cos(a)\!\left[ \sin(b z_m)x_m \! + \cos(b z_m)y_m \right]\! - \sin(a)z_m, \notag  \\
z_{m+1}\! &= \sin(a)\!\left[ \sin(b z_m)x_m \! + \cos(b z_m)y_m \right] \!+ \cos(a)z_m, \notag
\end{align}
which combines a rotation $R_x$ by angle $a$ around the $x$-axis and a torsion $S_z$ of amplitude $b$ around the $z$-axis~\footnote{See Supplementary Information (that contains Refs.~\cite{miller_1973,chirikov_1979,braun_1996,kus_1987}) for additional details on the kicked top, evolutions from different initial states, the effective Anderson Hamiltonian, quantum resonances, the random quantum kicked top and the Kullback-Leibler divergence.}.

\begin{figure}[t]
\begin{center}
\includegraphics[scale=1]{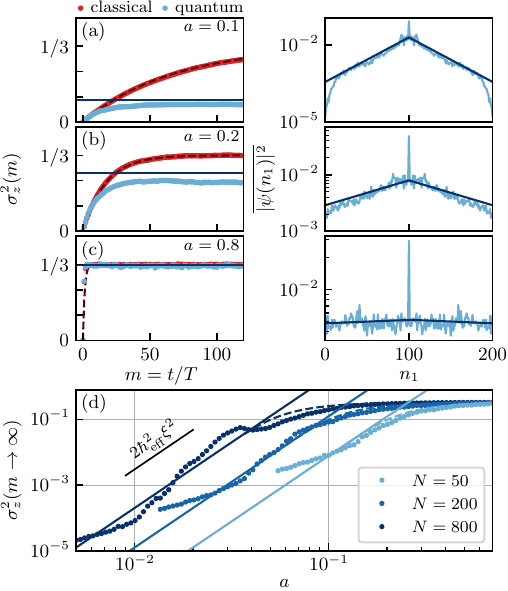}
\caption{
\textbf{Diffusion and localization.}
\textbf{(a)} Left: Variance of the population imbalance $\sigma_z^2(m)$ as a function of the number of modulation periods $m$ in the classical (light red, statistics based on $1000$ trajectories starting from $z_0=0$) and quantum (light blue, $N=200$ and $\ket{\psi_0}=\ket{100,100}$) regimes for $a=0.1$ and averaged over 100 realizations with $b \in [563,615]$. The classical exponential saturation of Eq.~\eqref{eq:exp_saturation} is plotted in dashed dark red, and the variance for an exponential localization (Eq.~\eqref{eq:exp_loc}) is plotted in solid dark blue. Right: Disorder-averaged Fock-space projection at final time $m=120$ (light blue) and exponential localization with $\xi=\sin^2(a)/(4\hbar_\deff^2)$ (dark blue).
\textbf{(b,c)} Same as (a) for $a=0.2$ and 0.8 respectively.
\textbf{(d)} Scaling of $\sigma_z^2(m\rightarrow\infty)$ as a function of $a$ for different atom numbers $N$ (disks, with $b\in[2340,2650]$), and variance of the exponentially localized state~\eqref{eq:exp_loc}, either truncated to the size of the Fock space (dashed lines) or not (solid lines).
}
\label{fig2}
\end{center}
\end{figure}

\paragraph{Bounded classical diffusion.} 
The map~\eqref{eq:map_kt} produces stroboscopic phase portraits that become increasingly chaotic as $b$ increases (for non-trivial rotations $a \neq 0 \mod \pi$), until chaotic trajectories predominate~\cite{nakamura_1986,haake_1987,kidd_2019} (see Fig. 1(c)). We focus on diffusion along the $z$-axis in this chaotic regime. From Eq.~\eqref{eq:map_kt}, the variance in $z$ for trajectories initialized at $z_0=0$ evolves as
\begin{equation}
\label{eq:exp_saturation}
    \sigma_{z}^2(m) \approx \dfrac{1}{3}\left[ 1-\left(1-\dfrac{1}{\tau}\right)^m \right]
    \overset{\tau \gg 1}{\approx} \dfrac{1}{3}\left(1-\de^{-m/\tau}\right),
\end{equation}
with the ergodization time \mbox{$\tau \equiv 2/(3\sin^2(a))$}~\cite{Note3}. As expected from ergodicity and the global chaoticity over the Bloch sphere (see Fig.~\ref{fig1}~(c)), $\sigma_{z,m}^2$ saturates to $1/3$ --- the variance of the uniform distribution over $[-1,1]$. This saturation is exponential in the regime $\tau \gg 1$, i.e. $|\sin(a)|\ll1$, for rotation angles $a$ close to 0 or $\pi$~\cite{fox_1994},
which reveals a short-time diffusive regime: $\sigma_{z}^2(m) \approx 2 D m$ for $m\ll\tau$, with the diffusion constant $D \equiv \sin^2(a)/4$. This behavior is confirmed in Fig.~\ref{fig2} for increasing values of $a$, where Eq.~\eqref{eq:exp_saturation} is seen to capture the classical dynamics accurately.

\paragraph{Quantum evolution and localization.} In the quantum regime, the $z$-axis of the Bloch sphere is spanned by the Fock states $\ket{n_1,n_2}$ (with $n_2 = N-n_1$), eigenstates of $\hJ_z$ (Fig.~\ref{fig1}~(a)). In this basis, the Floquet eigenvalue equation
$\hU_\dF \ket{\phi_j} = \de^{-i \varepsilon_j T/\hbar} \ket{\phi_j}$ can be cast into a 1D Anderson-like localization problem~\cite{anderson_1958,fishman_1982,grempel_1984,shepelyansky_1986,hainaut_2022} of finite size $L=N+1$, governed by the effective Hamiltonian
\begin{equation}
    \label{eq:AL_eq}
    \hH_\deff = \sum_{n=0}^NW_{j,n} |n\rangle\langle n| + \sum_{n,n'=0}^N t_{n',n} |n'\rangle\langle n|,
\end{equation}
where $W_{j,n} = \tan\left[ \varepsilon_j/2 - b(n-N/2)^2/N \right]$ acts as a quasi-random disorder potential, and the couplings $t_{n,n'}$ decay rapidly with $|n-n'|$~\cite{grempel_1984,hainaut_2022, Note3}. As we show, these features are sufficient to produce Anderson-like localization in the slow-diffusion regime.

The initial state $\ket{\psi_0}$ that we consider is the Fock state $\ket{N/2,N/2}$~\footnote{We consider an even number of particles for simplicity.}, which lies on the equator of the Bloch sphere and is thus the quantum analogue of the classical initial condition $z_0 = 0$, as seen from the Husimi representation~\cite{husimi_1940,agarwal_1981} in Fig.~\ref{fig1}~(d). Under iteration of the Floquet operator~\eqref{eq:U_kt}, the state spreads along $z$ until reaching a maximum extent, as seen in Figs.~\ref{fig1}~(d) and~\ref{fig2}. In the slow-diffusion regime $|\sin(a)|\ll1$, the quantum variance $\heff(\langle \hJ_z^2 \rangle - \langle \hJ_z \rangle^2)$ saturates well below its classical counterpart. This suppression of transport, here in Fock space, is the signature of Anderson localization~\cite{anderson_1958}.

The quantum evolutions presented here correspond to averages over many realizations of the effective disorder $W_{j,n}$ (Eq.~\eqref{eq:AL_eq}), achieved by sampling several values of the torsion parameter $b$. In doing so, two criteria must be satisfied: First $b$ must be large enough that the classical phase space remains globally chaotic~\cite{haake_1987,nakamura_1989} and $\sigma_z^2(m)$ follows Eq.~\eqref{eq:exp_saturation}. Second, $b$ must be sampled so as to avoid quantum resonances, which occur for $b=p/q \times \pi N$ (with integers $p$ and $q$) and artificially restore order in the dynamics~\cite{izrailev_1980,fishman_1982,Note3}.

The long-time Fock-space profiles in Fig.~\ref{fig2}~(a,b) are well described by the exponential decay
\begin{equation}
\label{eq:exp_loc}
    |\psi(n_1)|^2 \equiv |\langle n_1,n_2|\psi\rangle|^2 \sim \exp\left\{-|n_1-N/2|/\xi\right\},
\end{equation}
with localization length $\xi \approx D/\heff^2 = \sin^2(a)/(4\heff^2)$~\footnote{As $\sigma_z^2(m)$ depends on $z_0$, $\xi$ is expected to depend on $\langle \hJ_z \rangle_{\psi_0}$; see~\cite{Note3}.}, estimated from the localization time $t_\text{loc}/T \sim \xi$ required to reach the variance $\sigma^2_z = 2\heff^2\xi^2$ through early-time diffusion (Eq.~\eqref{eq:exp_saturation})~\cite{shepelyansky_1986,chirikov_1988,haake_1988}. Figure~\ref{fig2}~(c) illustrates the opposite regime of large rotation angles $a$, where $\tau \sim 1$, and the quantum variance saturates almost immediately, resulting in an ergodic-like, approximately uniform Fock-space profile with a ``localization length'' exceeding the system size $L=N+1$. Comparing these two scales yields the condition for Fock-space dynamical localization:
\begin{equation}
    \label{eq:localization_condition}
    \xi \ll L \Leftrightarrow |\sin(a)| \ll 4/\sqrt{N}.
\end{equation}
For any $a \neq 0 \mod\pi$, localization thus disappears as $N\rightarrow\infty$ ($\heff\rightarrow0$), in consistency with the classical ergodicity. An additional peak is observed in the center of the Fock-space profiles at long times (Fig.~\ref{fig2}~(a-c)). It corresponds to the enhanced return-to-the-origin probability~\cite{prigodin_1994,weaver_2000,engl_2014,hainaut_2017,hainaut_2018,engl_2014,dujardin_2016,schlagheck_2017,cherroret_2021, hummel_2022,karpiuk_2012, ghosh_2014, hainaut_2018, arrouas_2025}, and will be the subject of a future study.

Figure~\ref{fig2}~(d) shows the saturated variance $\sigma_z^2(m\rightarrow\infty)$ as a function of $a$ for $N=50$, 200 and 800, revealing three behaviors. For large $a$, condition~\eqref{eq:localization_condition} is not met and the system is ergodic, with $\sigma_z^2(m\rightarrow\infty)\approx 1/3$, matching the classical value. For smaller $a$, the state is exponentially localized, and $\sigma_z^2(m\rightarrow\infty) \approx 2 \heff^2\xi^2 = N^2 \sin^4(a)/32$, consistent with Eq.~\eqref{eq:exp_loc}. Finally, this scaling holds provided the localization length exceeds the spacing in Fock space, i.e. $\xi \gg 1 \Leftrightarrow |\sin(a)| \gg 4/N$~\cite{shepelyansky_1986}.

\begin{figure}[t]
\begin{center}
\includegraphics[scale=1]{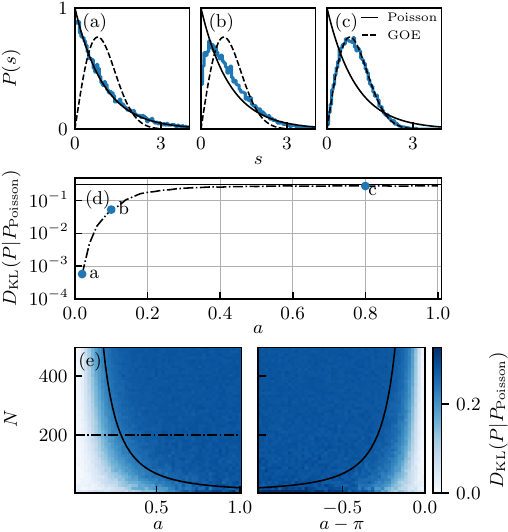}
\caption{
\textbf{Spectral statistics crossover towards localization} for the RQKT~\cite{Note3}. \textbf{(a,b,c)} Distribution of quasienergy spacings for $a=0.02$, 0.1 and  0.8 respectively for $N=200$ (solid blue lines) compared to RMT predictions (black lines), in linear scale. \textbf{(d)} Kullback-Leibler divergence $D_\dKL$ of the simulated spacing distributions to the Poisson distribution as a function of $a$ for $N=200$ (dash-dotted line) and identification of data (a-c) (blue disks). The KL divergence of GOE to Poisson is indicated with a solid black line. \textbf{(e)} Same $D_\dKL$ as a function of $N$ and $a$, with data (d) identified in dash-dotted line, and the coordinates of expected localization crossover $N = 16/\sin^2(a)$ (solid line, see Eq.~\eqref{eq:localization_condition}). All data is averaged over 100 realizations of the RQKT.
}
\label{fig3}
\end{center}
\end{figure}

\paragraph{Spectral statistics towards localization.} Localization can also be probed by the quasienergy-spacing statistics of the Floquet operator~\eqref{eq:U_kt}~\cite{sierant_2025}, as depicted in Fig.~\ref{fig3}. To fully avoid quantum resonances while systematically varying $N$, these results are obtained with the Random Quantum Kicked Top (RQKT), in which the torsion part of the evolution operator~\eqref{eq:U_kt} is replaced by random phases multiplying each Fock coefficient~\cite{altland_1996,Note3}. In the ergodic regime $|\sin(a)| \gg 4/\sqrt{N}$, the quasienergy levels exhibit linear repulsion (Fig.~\ref{fig3}~(c)), characteristic of chaotic systems, as modeled by the Gaussian Orthogonal Ensemble (GOE) from random-matrix theory~\cite{bohigas_1984}. Deep in the localized regime $|\sin(a)| \ll 4/\sqrt{N}$ (Fig.~\ref{fig3}~(a)), the statistics become Poissonian. At the crossover $\xi\sim L$ (Fig.~\ref{fig3}~(b)), a smooth interpolation between the two regimes is observed. This behavior can be quantified by the Kullback-Leibler divergence $D_\dKL$ of the simulated spacing distribution $P$ from the Poisson distribution $P_\text{Poisson}$~\cite{Note3}. Figure~\ref{fig3}~(d) and (e) display $D_\dKL$ as a function of $a$ and $N$, where the crossover $\xi\sim L$ is clearly observed, both for $a \ll 1$ and $|a-\pi| \ll 1$.

\paragraph{Connection to discrete time crystals.} Finally, the phenomenology of many-body localized discrete time crystals (DTCs) can be revisited from the MBDL perspective for $|a-\pi| \ll 1$~\cite{das_2025}. DTCs are periodically driven many-body systems that respond at a slower frequency than the drive, thereby breaking the discrete time-translation symmetry of the drive, in a manner robust to perturbations~\cite{sacha_2018,zaletel_2023}. A typical example is the driven spin-1/2 chain~\cite{yao_2017,choi_2017,zhang_2017}: without interactions, periodic $\pi$ pulses flip the entire chain, and the total magnetization oscillates at half the drive frequency, as two pulses are required to restore the initial configuration. While this fragile order breaks for any deviation of the pulse amplitude from $\pi$, interactions between the spins can stabilize this symmetry breaking~\cite{yao_2017,choi_2017,zhang_2017}. In our two-mode system, for $a = \pi$, the Floquet operator satisfies $\hU_\dF^2 = \mathds{1}$, so that the Floquet eigenstates pair into (anti)symmetric superpositions of Fock states separated by $\Delta n_1$, with an exact quasienergy difference $\Delta\varepsilon = h/2T$. For \mbox{$|a-\pi|\ll1$}, the $\pi$-pulse parity is broken, so that $\Delta\varepsilon$ is no longer pinned to $h/2T$. The DTC is nevertheless stabilized by the exponential localization of the Fock components of the superpositions, with localization length $\xi$, such that the quasienergy splitting is exponentially suppressed, \mbox{$|\Delta \varepsilon - h/2T| \sim \de^{-\Delta n_1/2\xi}$}~\cite{doggen_2017,hebraud_2026}.

\begin{figure}[t]
\begin{center}
\includegraphics[scale=1]{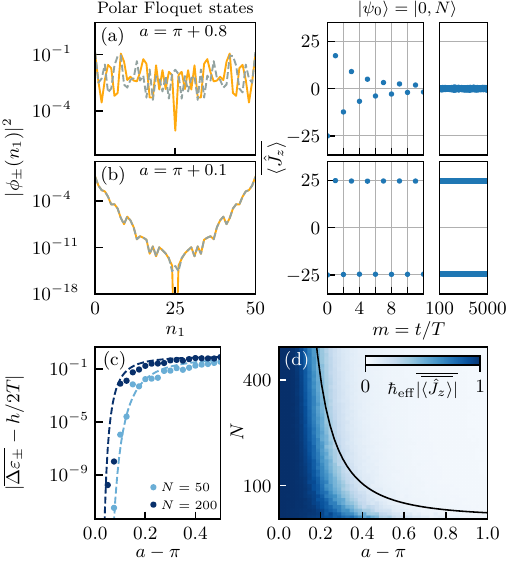}
\caption{
\textbf{Discrete time crystals from the perspective of MBDL.} \textbf{(a)} Left panel: Fock space projection of the two Floquet states with maximal overlap onto $\ket{0,N}$ and $\ket{N,0}$ for $a = \pi + 0.8$. Right panels: Expectation value of the population imbalance $\langle \hJ_z \rangle$ as a function of $m$ for $\ket{\psi_0} = \ket{0,N}$.
\textbf{(b)} Same as (a) for $a=\pi+0.1$.
\textbf{(c)} Degeneracy lifting $|\Delta \varepsilon_\pm - h/2T|$ as a function of $a-\pi$ for $N=50$ (light blue) and $N=200$ (dark blue), with numerics (dots) and the estimate $\de^{-N/2\xi}$ (dashed lines)~\cite{Note5}.
\textbf{(d)} Time average of $|{\langle\hJ_z\rangle}|$ over 200 modulation periods for $\ket{\psi_0}=\ket{0,N}$ as a function of $a$ and $N$, with localization crossover $N = 16/\sin^2(a)$ (solid line, see Eq.~\eqref{eq:localization_condition}).
Results obtained on the RQKT, with an average over 100 realizations for all panels except $|\phi_\pm(n_1)|^2$.
}
\label{fig4}
\end{center}
\end{figure}

In particular, a fully polarized initial state $\ket{\psi_0} = \ket{0,N}$ projects predominantly onto Floquet eigenstates $\ket{\phi_\pm}$ located near the poles of the Bloch sphere (with $\Delta n_1=N$), as shown in Fig.~\ref{fig4}~(a,b). The resulting quasienergy splitting $|\Delta\varepsilon_\pm - h/2T| \sim \de^{-N/2\xi}$ (Fig.~\ref{fig4}~(c)) gives rise to period-$2T$ oscillations persisting for exponentially long time (Fig.~\ref{fig4}~(b)). Figure~\ref{fig4}~(d) shows the time-averaged $|\langle \hJ_z \rangle|$ over 200 periods as a function of $a$ and $N$, which acts as a DTC order parameter, remaining large only when the oscillation at period 2T persists with near-maximal amplitude --- below the localization crossover~\eqref{eq:localization_condition}.
The robustness of this DTC is thus a direct consequence of the dynamical localization of the Floquet eigenstates, which prevents $\ket{\psi_0}$ from propagating away from the edges of the $z$-axis~\footnote{The exact rotational symmetry $e^{i\pi\hat{J}_x}\hat{H}(t)e^{-i\pi\hat{J}_x} = \hat{H}(t)$ of the Hamiltonian~\eqref{eq:hkt}, 
preserved for all $a$, further contributes to this robustness by ensuring that 
the Floquet eigenstates retain a clean pairing structure. We indeed observe that 
explicitly breaking this symmetry, without affecting the localization, accelerates 
the melting of the DTC; see~\cite{Note3}.}.

\paragraph{Discussion and conclusion.} We have revealed the existence of a dynamically localized regime for interacting bosons in a kicked two-mode system --- a model exactly mapped to the quantum kicked top. This MBDL has largely eluded previous studies of the kicked top, whether in its spin or many-body interpretation, as the coupling parameter $a$ has typically been fixed near $\pi/2$ in the literature~\cite{nakamura_1986,haake_1987,kus_1988,dietz_1991,saher_1991,d_ariano_1992,kus_1993,schack_1994,fox_1994,peres_1996,schack_1996,lerose_2020}, focusing on the ergodic regime of Fig.~\ref{fig2}~(c), where the underlying classical chaos was shown to drive entropy growth~\cite{peres_1994,alicki_1996,miller_1999,zyczkowski_2001,tanaka_2002,weinstein_2002,bandyopadhyay_2004,wang_2004,ghose_2004,demkowicz_2004,xie_2005,weinstein_2006,stamatiou_2007,ghose_2008,chaudhury_2009,neill_2016,pappalardi_2018,piga_2019,kumari_2019,herrman_2019,lerose_2020,li_2021,wang_2023} and information scrambling~\cite{swingle_2016,pappalardi_2018,lerose_2020,omanakuttan_2023} at the quantum level.

Our results share key signatures with MBL: interference-induced localization of the many-body state, memory of initial conditions, ergodicity breaking of the corresponding classical dynamics and emergent integrability as probed by spectral statistics --- further supported by the direct connection to MBL discrete time crystals. We note, however, that localization here occurs in Fock space (i.e. onto the eigenbasis of the interaction operator) and in a driven setting, so that strong effective disorder corresponds counterintuitively to small kick amplitude $a$, the regime where interactions dominate over the kicking and where localization becomes naturally expected. Moreover, whether an MBL phase survives in the thermodynamic limit remains an open question, not addressed by this minimal system.

A natural extension of this work is to apply the Fock-space localization picture to driven bosonic systems with more modes~\cite{vanhaele_2022,creffield_2025}. In particular, since a fully connected $K$-mode bosonic system maps to a $(K-1)$-dimensional Fock space, we expect the existence of an Anderson localization transition for $K\geq4$ all-to-all connected modes~\cite{chabe_2008,olsen_2025,olsen_2025b}.

The Fock-space localization described here could be observed experimentally with cold-atom ensembles in two-mode systems --- where both the mean-field~\cite{anker_2005,albiez_2005,ryu_2013,ji_2022} and quantum~\cite{sebby_strabley_2007,folling_2007,esteve_2008,gross_2010,maussang_2010,fadel_2018,zhang_2024,imperto_2024} regimes have been realized --- starting from a fully polarized state $\ket{\psi_0} = \ket{N,0}$~\cite{Note3}, which is easily prepared in practice. Localization would then manifest as an exponential decay of the long-time population-imbalance distribution around $\langle \hJ_z \rangle_{\psi_0}$. Alternatively, this physics could be emulated with a large quantum spin, as realized in spinor gases~\cite{chaudhury_2009,evrard_2019} or silicon nanoelectronic devices~\cite{fernandez_de_fuentes_2024}, where $\hR_x$ and $\hS_z$ of Eq.~\eqref{eq:U_kt} can be engineered via electromagnetic fields.

\paragraph{Acknowledgements.} We thank Felix Palm for discussions. This work was supported by the ERC Grant LATIS, the EOS Project CHEQS, the ANR PEPR Grants QUTISYM ANR-23-PETQ-0002 and Dyn1D ANR-23-PETQ-0001, and partly by the ANR Research Grant ManyBodyNet ANR-24-CE30-5851. 

\bibliography{bibliography}

\clearpage

\onecolumngrid
\setcounter{equation}{0}
\setcounter{figure}{0}

\renewcommand{\theequation}{S\arabic{equation}}
\renewcommand{\thefigure}{S\arabic{figure}}
\renewcommand{\thetable}{S\arabic{table}}

\begin{center}
    {\large\bfseries Supplementary Material to ``Many-body dynamical localization in Fock space''}
\end{center}



\section{I. Classical diffusion of the kicked top and evolutions from different initial states}
\label{app:classical_dyn}
The classical map for the kicked top (Eq.~4 for the main text) is derived from the Floquet operator 
\begin{equation}
\label{eq:Uf}
\hU_\dF = \de^{-\di\hbar_\deff b \hJ_z^2}\de^{-\di a \hJ_x},
\end{equation}
by evaluating the stroboscopic evolution of the pseudo-spin operators in the Heisenberg picture: $\hJ_\mu(m+1) = \hU_\dF^\dagger \hJ_\mu(m) \hU_\dF$ (with $\mu=x,y,z$). This can be carried out explicitly using the Baker-Campbell-Hausdorff identity $\de^AB\de^{-A} = \sum_{k=0}^\infty [A,B]_k/k!$~\cite{miller_1973}, where $[A,B]_k = [\underbrace{A,...[A,[A}_{k\;\text{times}},B]]...]$ denotes the nested commutators, together with the $SU(2)$ commutation relations $[\hJ_\mu,\hJ_\nu] = \di \varepsilon_{\mu\nu\lambda} \hJ_\lambda$. The classical limit follows by taking $N\rightarrow\infty,\heff\rightarrow0$, so that the commutators $[\hat{\mu},\hat{\nu}] = \di\heff \varepsilon_{\mu\nu\lambda}\hat{\lambda}$ between the rescaled pseudo-spins $\hat{\mu} = \heff \hJ_\mu$ vanish, allowing one to describe the state of the system by the classical phase-space coordinates $(x,y,z)$. The resulting map is
\begin{align}
\label{eq:map_kt2}
x_{m+1}\! &= \cos(b z_m) x_m - \sin(b z_m) y_m, \\
y_{m+1}\! &= \cos(a)\!\left[ \sin(b z_m)x_m \! + \cos(b z_m)y_m \right]\! - \sin(a)z_m, \notag  \\
z_{m+1}\! &= \sin(a)\!\left[ \sin(b z_m)x_m \! + \cos(b z_m)y_m \right] \!+ \cos(a)z_m, \notag
\end{align}
which is Eq.~(4) of the main text. The apparent inversion in the order of rotation and torsion between Eqs.~\eqref{eq:Uf} and~\eqref{eq:map_kt2} is a consequence of the Heisenberg picture, in which operators rather than states evolve. The two equations are nonetheless consistent in the stroboscopic picture.

\begin{figure}[t]
\begin{center}
\includegraphics[scale=1]{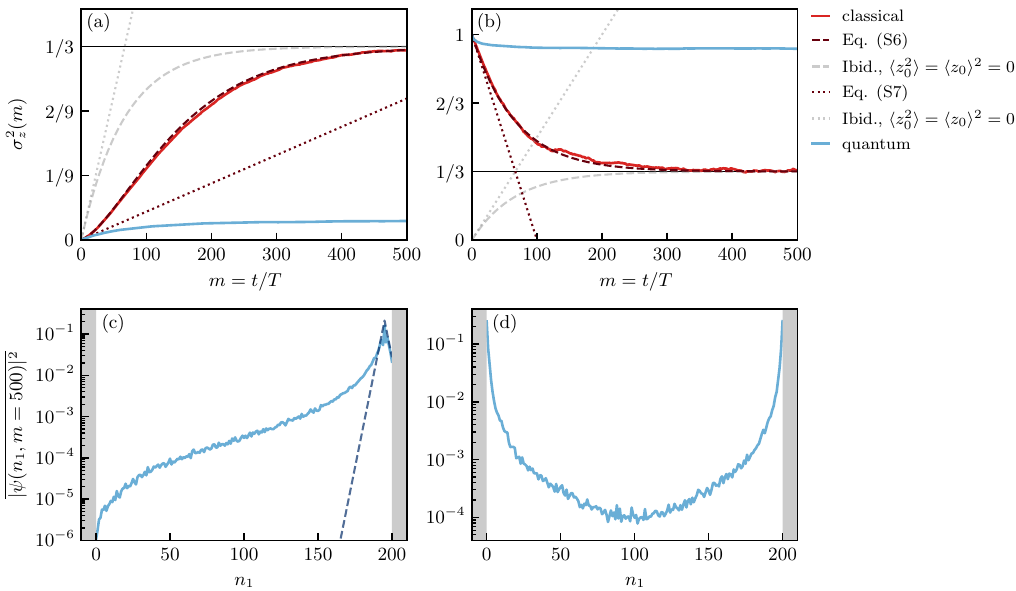}
\caption{
\textbf{Diffusion and localization from different initial states.}
\textbf{(a)} Variance of the population imbalance $\sigma_z^2(m)$ as a function of the number of modulation periods $m$, for $a=0.1$. Classical results (solid light-red curve): average over 500 trajectories starting at $z_0= 0.95$, further averaged over 100 realizations with $b \in [600, 700]$. Quantum results (solid light-blue curve): $N = 200$, $\ket{\psi_0}=\ket{195,5}$ (i.e. $n_1(0) = N(z_0+1)/2$), evolved with the unsymmetrized RQKT 
(see Sec.~III) 
and averaged over 100 realizations. Dashed curves show the analytical prediction of Eq.~\eqref{eq:varz0}. Dotted curves show the early-time diffusion Eq.~\eqref{eq:diffusion_z0}. Both are shown in dark red for $\langle z_0^2 \rangle = \langle z_0 \rangle^2 = (0.95)^2$ and gray for the equatorial case $\langle z_0^2 \rangle$ = $\langle z_0 \rangle^2 = 0$ (as in Figs.~1 and 2 of the main text).
\textbf{(b)} Same as (a) for initial conditions at the poles, $\{ (x,y,z)_0\} = \{(0,0,-1),(0,0,1)\}$, with the corresponding quantum initial condition $\ket{\psi_0} = (\ket{200,0}+\ket{0,200})/\sqrt{2}$.
\textbf{(c)} Disorder-averaged Fock-space projection at $m=500$ (solid light-blue curve; 100 realizations), compared to the exponential localization profile of Eq.~(7) of the main text with $\xi = D/\heff^2$ (dashed dark-blue curve), with $D$ estimated from Eq.~\eqref{eq:Dz0}. The gray stripes indicate the edges of the Fock space.
\textbf{(d)} Same as (c) for $\ket{\psi_0} = (\ket{200,0}+\ket{0,200})/\sqrt{2}$. In this case $D<0$, so the localization length estimated from the diffusion coefficient~\eqref{eq:Dz0} does not characterize the localization scale.
}
\label{fig_different_z0}
\end{center}
\end{figure}

Since $x_m^2+y_m^2+z_m^2=1$, the $z$-coordinate evolves according to
\begin{equation}
    \label{eq:zp_z}
    z_{m+1} = \cos(a) z_m + \sin(a) \sqrt{1-z_m^2} \sin(\varphi_m + bz_m),
\end{equation}
where $\varphi_m+b z_m$ is the evolution of the azimuthal angle $\varphi_m = \arg(x_m+\di y_m)$ generated by the torsion. In the chaotic regime associated with large values of $b$, one can make the approximation that these angles are not correlated, and thus behave as random variables uniformly distributed over $[0,2\pi]$ when averaging over many initial conditions~\cite{chirikov_1979} (an operation denoted $\langle \cdot\rangle$ below~\footnote{Numerically, we further average over values $b$, which can be required to average over different realizations depending on the initial condition considered (e.g. at the poles in $|z_0|=1$, where all azimuths collapse onto a unique point producing the same trajectory).}). Under this assumption, Eq.~\eqref{eq:zp_z} yields the recursion relations
\begin{equation}
    \label{eq:zm}
    \begin{split}
        \langle z_{m+1} \rangle  &\approx \cos(a) \langle z_m \rangle\\
        \langle z_{m+1}^2 \rangle  &\approx \cos^2(a) \langle z_m^2 \rangle + \dfrac{1}{2}\sin^2(a)\left(1-\langle z_m^2 \rangle\right),
    \end{split}    
\end{equation}
and by iteration
\begin{equation}
    \label{eq:z2m}
    \begin{split}
        \langle z_m \rangle &\approx \cos^m(a) \langle z_0 \rangle,\\
        \langle z_m^2 \rangle &\approx \left(1-\dfrac{3}{2}\sin^2(a)\right)^m\left( \langle z_0^2 \rangle - \dfrac{1}{3}\right) + \dfrac{1}{3}.
    \end{split}
\end{equation}
The variance then follows directly as
\begin{equation}
\label{eq:varz0}
    \sigma_{z}^2(m) \approx \left(1-\dfrac{3}{2}\sin^2(a)\right)^m \left(\langle z_0^2 \rangle - \dfrac{1}{3}\right) - \cos^{2m}(a) \langle z_0 \rangle^2 + \dfrac{1}{3},
\end{equation}
with $\sigma_z^2(m\rightarrow\infty) = 1/3$ if $a \neq 0 \mod \pi$, and $\sigma_z^2(m\rightarrow\infty) = \sigma_z^2(0) = \langle z_0^2 \rangle - \langle z_0\rangle^2$ otherwise. In the general context of Eq.~\eqref{eq:varz0} and for small values of $a$, a short-time diffusion is also observed (see main text):
\begin{equation}
    \label{eq:diffusion_z0}
    \sigma_z^2(m) \approx \sigma_z^2(0) + 2D(\langle z_0^2\rangle,\langle z_0 \rangle ^2)m,
\end{equation}
with 
\begin{equation}
\label{eq:Dz0}
    D(\langle z_0^2\rangle,\langle z_0 \rangle ^2) = \dfrac{\sin^2(a)}{4}\left(1-3\langle z_0^2\rangle + 2\langle z_0\rangle^2\right).
\end{equation}
Notably, this diffusion coefficient is negative when $\sigma_z^2(0) > (1-\langle z_0\rangle^2)/3$. This occurs, for instance, when starting from the two poles of the Bloch sphere $\lbrace (x,y,z)_0 \rbrace = \lbrace (0,0,1),(0,0,-1) \rbrace$, for which the initial variance is maximal ($\sigma_z^2(0)=1$), and decays toward $1/3$ (exponentially so for $|\sin(a)| \ll 1$). In the special case $\langle z_0 \rangle^2 = \langle z_0^2 \rangle = 0$, both Eqs.~\eqref{eq:varz0} and~\eqref{eq:Dz0} reduce to the results of the main text.

Diffusion from initial states other than the equatorial configuration studied in the main text is presented in Fig.~\ref{fig_different_z0}. Figure~\ref{fig_different_z0}~(a) shows the case $z_0=0.95$, where classical trajectories starting from different azimuths $\varphi_0=\arg(x_0+\di y_0)$ are compared with the disorder-averaged evolution of the corresponding quantum initial state $\ket{\psi_0}=\ket{195,5}$ (with $z = (n_1-n_2)/N$). Figure~\ref{fig_different_z0}~(b) shows the case of trajectories starting at the poles of the Bloch sphere, $(x,y,z)_0 = (0,0,\pm1)$, for which the diffusion coefficient is negative, $D=-\sin^2(a)/2$. The corresponding quantum initial state is the NOON state $\ket{\psi_0}=(\ket{200,0}+\ket{0,200})/\sqrt{2}$. In both cases, Eq.~\eqref{eq:varz0} accurately captures the classical dynamics. Figure~\ref{fig_different_z0}~(c) and (d) show the Fock-space populations at long time, where the expected dependence of the localization length on $z_0$ is observed.

From Eq.~\eqref{eq:zp_z}, the average scattering range in $z$ for trajectories starting at $z_0$ can be estimated as
\begin{equation}
\label{eq:hopping_range}
    \ell(z_0) \approx \sqrt{ \left\langle (z_1-z_0)^2 \right\rangle} \approx \sqrt{\left[\cos(a)-1\right]^2 z_0^2 + \sin^2(a)\left(1-z_0^2\right)/2}.
\end{equation}

\section{II. Effective Anderson Hamiltonian}
\label{app:H_eff}

The effective 1D Anderson Hamiltonian (Eq.~(6) of the main text) is obtained by following Refs.~\cite{fishman_1982,grempel_1984,shepelyansky_1986}. We start with the Floquet states $\ket{\phi_j^-}$ (resp. $\ket{\phi_j^+}$), eigenstates of the periodic evolution just before (resp. after) the rotation $\hR_x(a) = \de^{-\di a \hJ_x}$. Note that, in the main text, $\ket{\phi_j} \equiv \ket{\phi_j^-}$. The states $\ket{\phi_j^\pm}$ share the same quasienergy $\varepsilon_j$:
\begin{equation}
\label{eq:phi_pm}
\begin{split}
    \hS_z(b)\hR_x(a) \ket{\phi_j^-} &= \de^{-\di \varepsilon_j T/\hbar} \ket{\phi_j^-},\\
    \hR_x(a)\hS_z(b) \ket{\phi_j^+} &= \de^{-\di \varepsilon_j T/\hbar} \ket{\phi_j^+},\\
    \ket{\phi_j^+} &= \hR_x(a) \ket{\phi_j^-},
\end{split}
\end{equation}
These states are expanded over the Fock basis $\{\ket{n_1}\equiv\ket{n_1,n_2 = N-n_1}\}$ as \mbox{$\ket{\phi_j^\pm} = \sum_{n=0}^N c^\pm_n \ket{n}$} (omitting the label 1). From Eqs.~\eqref{eq:phi_pm}, $c_n^\pm$ can be expressed from one another as
\begin{equation}
\label{eq:cn_pm}
c_n^- = \de^{\di (\varepsilon_j - \heff b (n-N/2)^2)} c_n^+.
\end{equation}
We define the average eigenstate~\cite{fishman_1982}
\begin{equation}
    \ket{\Phi_j} \equiv (\ket{\phi_j^+}+\ket{\phi_j^-})/2 = \sum_n c_n \ket{n},
\end{equation}
with Fock coefficients $c_n = (c_n^- + c_n^+)/2$, such that
\begin{equation}
\label{eq:cn_vs_W}
    c_n^\pm = c_n \pm \di \sum_{n'=0}^{N} t_{n,n'} c_{n'},
\end{equation}
with
\begin{equation}
\label{eq:tnn}
t_{n,n'} = \langle n | \tan(-a\hJ_x/2) |n'\rangle.
\end{equation}
Since $\hJ_x = \sum_{k_x=0}^N (k_x - N/2) |k_x\rangle\langle k_x|$, with $\langle n|k_x \rangle = \langle n | \de^{-\di \pi \hJ_y /2} | k \rangle$, these couplings can be written as:
\begin{equation}
\label{eq:tnn2}
    t_{n,n'} = \sum_{k=0}^{N} \tan\left[-\dfrac{a}{2}\left(k-\dfrac{N}{2}\right)\right] d^{(y)}_{n,k}\left(\dfrac{\pi}{2}\right)d^{(y)}_{n',k}\left(\dfrac{\pi}{2}\right),
\end{equation}
where $d^{(y)}_{n,k}(\theta)$ is the Wigner $d$-matrix~\cite{fox_1994,braun_1996}:
\begin{equation}
\label{eq:d_matrix}
\begin{split}
    d^{(y)}_{n,k}(\theta) &= \langle n | \de^{-\di \theta \hJ_y} | k \rangle\\
    &= \sqrt{n!\,(N-n)!\,k!\,(N-k!)}\sum_{s=s_\dmin}^{s_\dmax} \dfrac{(-1)^{n-k+s} \left(\cos \frac{\theta}{2} \right)^{N+k-n-2s} \left( \sin \frac{\theta}{2}\right)^{n-k+2s}}{(k-s)!\,s!\,(n-k+s)!\,(N-n-s)!}.
\end{split}
\end{equation}
In Eq.~\eqref{eq:d_matrix}, the summation indices $s_\dmin = \dmax\{0, k-n\}$ and $s_\dmax = \dmin\{k,N-n\}$ are such that the factorial arguments remain nonnegative. Unlike the kicked rotor, the couplings $t_{n,n'}$ of Eq.~\eqref{eq:tnn} are not translationally invariant, a property that could be used to study Anderson localization in curved spacetime~\cite{li_2023}. In the slow-diffusion regime (see main text), their range is approximately given by the classical scattering range of Eq.~\eqref{eq:hopping_range}. Combining Eqs.~\eqref{eq:cn_pm} and~\eqref{eq:cn_vs_W} finally gives
\begin{equation}
\label{eq:H_Anderson_Fock}
        W_{j,n} c_n + \sum_{n,n'} t_{n,n'} c_{n'} = 0,
\end{equation}
with
\begin{equation}
\label{eq:w_jn}
        W_{j,n} = \tan{\left[ \varepsilon_j/2-b(n-N/2)^2/N \right]}.
\end{equation}
Equation~\eqref{eq:H_Anderson_Fock} has the form of a stationary Schrödinger equation for the Anderson tight-binding Hamiltonian~(6) of the main text, with $\ket{\Phi_j}$ its zero-energy eigenstate for each $\varepsilon_j$. Equation~\eqref{eq:w_jn} shows that different values of $b$ produce different realizations of the effective disorder.

\section{III. Quantum resonances and the random quantum kicked top}
\label{app:RQKT}
\begin{figure}[t]
\begin{center}
\includegraphics[scale=1]{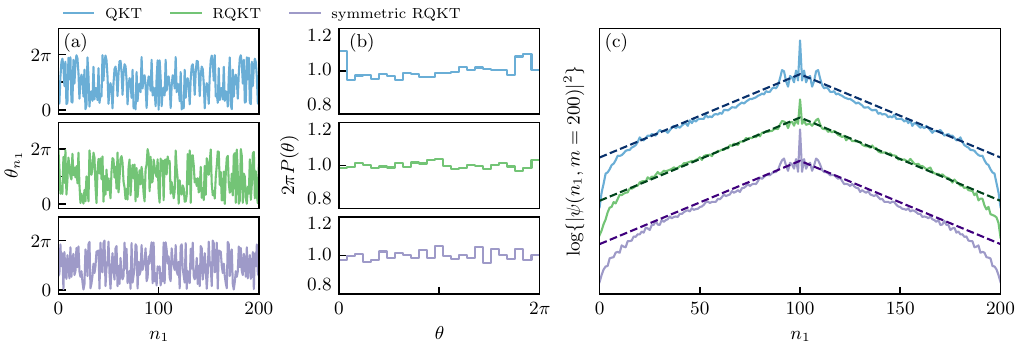}
\caption{
\textbf{Random quantum kicked top} (RQKT).
\textbf{(a)} Top: Phases $\theta_{n_1}$ acquired by the Fock coefficient $\psi(n_1) = \langle n_1|\psi\rangle$ under the torsion operator $\hS_z(b)$, for three models: the standard quantum kicked top (QKT; Eq.~(2) of the main text; drawn in blue for $b=563$), the RQKT (green) and the symmetric RQKT (purple, see text).
\textbf{(b)} Distribution of $\theta_{n_1}$ for 500 realizations for each model. For the QKT (blue), $b$ is sampled uniformly over $[563,615]$ (as in Fig.~1 of the main text); the slight overrepresentation of phase values around $\theta_{n_1} = 0 \mod 2\pi$ stems from a few high-order quantum resonances encountered in this interval.
\textbf{(c)} Fock-space population $|\psi(n_1)|^2$ after $m=200$ modulation periods (solid colored lines), for the initial state $\ket{\psi_0} = \ket{N/2,N/2}$ evolved under each of the three models with the kick amplitude $a = 0.1$, yielding the same exponential localization --- Eq.~7 of the main text is drawn in dashed lines, replicated three times. Profiles are offset vertically for clarity, with QKT, RQKT and symmetric RQKT from top to bottom. $N=200$ for all panels.
}
\label{fig_rqkt}
\end{center}
\end{figure}

Figures~3 and 4 of the main text present results obtained with the random quantum kicked top (RQKT), in which the torsion operator $\hS_z(b) = \de^{\di\heff b\hJ_z^2}$ is replaced by random phases multiplying each Fock coefficient (an approach commonly employed to model the quantum kicked rotor~\cite{altland_1996}). The RQKT is introduced in order to avoid quantum resonances~\cite{izrailev_1980,fishman_1982}, which suppress the effective disorder produced by the kicking (see Eqs.~\eqref{eq:H_Anderson_Fock} and \eqref{eq:w_jn}). Indeed, the torsion operator is diagonal over the Fock basis, with matrix elements $\langle n'_1 | \hS_z(b) |n_1 \rangle = \de^{-\di \theta_{n_1}} \delta_{n'_1,n_1}$, where $\theta_{n_1} = (2b/N)(n_1-N/2)^2$. When $b = p/q \times \pi N$ for integers $p,q$, these phases become rational multiples of $\pi$, thereby restoring order in the system (a dramatic instance being $b=\pi N$, where the torsion reduces to the identity). For a given $N$, such resonances can be avoided by sampling $b$ away from these special values, as done in Figs.~1 and 2 of the main text. In the regime $b \gg \pi N$,
the phases $\theta_{n_1}$ are well approximated by uniformly distributed random variables on $[0,2\pi]$, yielding essentially identical results, as illustrated in Fig.~\ref{fig_rqkt}.

In our study, the Floquet operator possesses the rotational symmetry $\hU_\dF = \hR_x(\pi)^\dagger \hU_\dF \hR_x(\pi)$, which partitions the Hilbert space into two subspaces of opposite parity. This symmetry is generically broken by the RQKT, but can be restored by symmetrizing the random phases $\theta^{sym}_{n_1} \equiv \theta_{n_1} + \theta_{N-n_1} \mod 2\pi$. In Fig.~3 of the main text, the phases are left unsymmetrized, since the absence of Hilbert space partition simplifies the spectral analysis (without loss of generality for our purposes)~\cite{haake_1987,kus_1987,hebraud_2026}. For the results of Fig.~4, however, the rotational symmetry is enforced. Indeed, we observe that breaking this symmetry accelerates 
the melting of the DTC, as an initial Fock state then projects onto eigenstates with no 
clean pairing structure, resulting in a greater variety of quasienergy 
splittings contributing to $\langle \hJ_z \rangle_{\psi(t)}$, and thus 
faster dephasing upon averaging over disorder realizations.

\section{IV. Kullback-Leibler divergence}
\label{app:kl_div}
In Fig.~3 of the main text, we employ the Kullback-Leibler divergence $D_\dKL$ to quantify the distance between the quasienergy-difference distribution $P(s)$ obtained numerically and the Poisson distribution $P_\text{Poisson} = \de^{-s}$, expected for integrable systems. This quantity is defined as
\begin{equation}
    D_\dKL(P|P_\text{Poisson}) = \int_0^{s_\dmax} P(s) \ln\left(\dfrac{P(s)}{P_\text{Poisson}(s)}\right)\dd s,
\end{equation}
where $s_\dmax$ is a cutoff introduced to ensure that the numerical values $P(s\leq s_\dmax) > 0$. We used $s_\dmax = 2$ for the data of Fig.~3.

\end{document}